# The two subset recurrent property of Markov chains

Lars Holden, Norsk Regnesentral

## Abstract

This paper proposes a new type of recurrence where we divide the Markov chains into intervals that start when the chain enters into a subset $A$, then sample another subset $B$ far away from $A$ and end when the chain again return to $A$. The length of these intervals have the same distribution and if $A$ and $B$ are far apart, almost independent of each other. $A$ and $B$ may be any subsets of the state space that are far apart of each other and such that the movement between the subsets is repeated several times in a long Markov chain. The expected length of the intervals is used in a function that describes the mixing properties of the chain and improves our understanding of Markov chains.

The paper proves a theorem that gives a bound on the variance of the estimate for $\pi(A)$, the probability for $A$ under the limiting density of the Markov chain. This may be used to find the length of the Markov chain that is needed to explore the state space sufficiently. It is shown that the length of the periods between each time $A$ is entered by the Markov chain, has a heavy tailed distribution. This increases the upper bound for the variance of the estimate $\pi(A)$.

The paper gives a general guideline on how to find the optimal scaling of parameters in the Metropolis-Hastings simulation algorithm that implicit determine the acceptance rate. We find examples where it is optimal to have a much smaller acceptance rate than what is generally recommended in the literature and also examples where the optimal acceptance rate vanishes in the limit.

## Introduction

In Markov chains a state is recurrent if the probability is 1 for coming back to the same state. The recurrence periods between each time the state is visited are identically, independently distributed (Meyn and Tweedie, 1993).  For continuous state Markov chains it is more useful to define a subset $A$ such that a long chain visits the subset many times. In order to utilize a repetition, we define the start of a new period each time the chain enters $A$. Since we are not interested in short periods where the chain moves at the border of the subset, we define another subset $B$ far apart from $A$ and require that $B$ is visited in each period. Each period, denoted recurrence interval, have length from the same distribution and the lengths will be almost independent. Then the correlation between lengths from different recurrence intervals is negligible. Analyzing these lengths improves the understanding of the mixing of the Markov chain, finds optimal simulation parameters and bounds the variance of the estimate for the probability $\pi(A)$.  We focus on stationary Markov chains that define a limiting density $\pi(.)$, where the probability for a subset in the limiting density is equal to the probability that a state is inside the subset.

In Section two we define recurrence intervals. Here we also propose a function of the expected length of the recurrence intervals and the probability of the subset $A$. This function describes the mixing of the Markov chain. When the subset becomes more extreme, then the length of the recurrence intervals increases. The recurrence interval is related to the path between states in a finite state space defined in Diaconis and Stroock, 1991. By using the Poincaré inequality this was used as a bound for the second largest eigenvalue that is important for the convergence of the Markov chain.

In Section three, the recurrence intervals are used to give a bound on the variance of an estimate of the probability of $A$ for the limiting density. The bound depends on the expected length of the recurrence intervals but also the distribution of the length. The authors experience is that the length typically has a heavy tailed distribution, but this depends of course on the Markov chain and the states $A$ and $B$. In Section six we make a statistical analysis of the recurrence intervals in four models. The length of the recurrence intervals is modelled with a Weibull distribution. Except for an "extreme" model, the shape parameter $k > 1$, giving a distribution with lighter tail than the exponential distribution. The ratio between the fraction of states inside $A$ in the recurrence interval and the probability for $A$ is best modelled with the exponential distribution with parameter 1.

In this paper we use the Metropolis-Hastings algorithm, but the results are valid for more general Markov chains. The Metropolis-Hastings algorithm generates an ergodic Markov chain $\{x_i\}$ converging to a target density $\pi(x)$. In each iteration $i$, a new state $y_{i+1}$ is proposed depending on the previous state $x_i$ according to a proposal density $q(y_{i+1}; x_i)$. We focus in this paper on symmetric random walk Metropolis-Hastings where $y_i = x_i + z_i$ and $z_i$ is i.i.d from a symmetric distribution. This means that $q(y; x) = q(x; y)$. The acceptance rate is the fraction of iterations when $x_{i+1} = y_{i+1}$.

Previously, it is proved under quite general assumptions that we obtain the fastest convergence of the Metropolis-Hastings algorithm when the acceptance rate is 0.234. Hence, it has become customary to scale the proposal function in order to obtain an acceptance rate close to this number quite generally also in the cases where we don't know whether this is optimal or not. In Section four we give a general advice on how to find the optimal acceptance rates based on recurrence intervals. In Section five we analyze the chain between A and B. We illustrate some problems in two toy examples and from a large complex climate model.

There are alternatives to Markov chains where it is possible increase convergence and mixing by use of adaption. This is a huge area with many promising techniques presented by e.g. Giordani and Kohn 2006. Other alternatives are state dependent scaling, jumping between parallel chains or Langevin (MALA) algorithms, see Roberts and Rosenthal (1998). However, many problems may not use these techniques particularly in large complex models. For many non-linear processes Markov chain is the only alternative. These problems are typically very computer intensive and it is necessary to tune the simulation parameters in order to make the algorithm as efficient as possible. Still, it may only be possible to run one or a few chains within the available computer and time resources making it necessary to evaluate convergence from one or a few chains. This paper focuses on problems where it is necessary to use Markov chains. Markov chain is still an active research area, e.g. Rosenthal and Rosenthal 2015.

## 2. Mixing of Markov chains and recurrence periods.

The two most important properties of Markov chains are convergence, i.e. how fast the chain converges to the target distribution from an initial state and mixing i.e. how fast element $x_{i+d}$ in the chain becomes independent of the previous element $x_i$. Convergence is important because it allows us to determine the length of the burn in period and mixing is important to estimate how many elements in the chain are necessary in order to get at good representation of the limiting distribution. Both these properties are connected to how fast the norm $sup_A |P^n(x,A) - \pi(A)|$ vanishes when $n$ increases. Here $P^n(x,A)$ is the n-step Markov chain kernel, i.e. the probability for $x_n \in A$ when starting in initial state, i.e. $x = x_0$. It is usual to use the total variation norm, but also other norms are used and of interest. For convergence we are interested in the expression for $x = x_0$, the start position of the chain. For mixing, the expression is important for all values of $x$ in the state space. In this paper we are focusing on mixing.

There are few papers on mixing of Markov chains and no established quantitative measures. This is in contrast convergence where there are many papers on convergence and several possible norms are discussed, e.g. Meyn and Tweedie, (1993) and Holden, (1998). It is usual to check the autocorrelation and other convergence characteristics, see f.ex. Cowles and Carlin, (1996). However, they also conclude that there are few methods of practical use and these do not identify or solve all problems. For example minimizing autocorrelation may result in small step length that makes it very difficult to find other modes. Then the number of iterations needed in order to sample the target density properly increases dramatically. The authors experience based on many years of using Markov chains is that one should use many different measures in order to understand the chain and the limiting distribution as good as possible. This is discussed in several papers e.g. (Rosenthal, 2010) and in other parts of this Handbook. Regeneration is another approach, see Roberts and Tweedie, (1999).

It is not easy to quantify numerically the total variation or most other used norms. Therefore, these norms are not well suited for comparing the convergence and mixing of different proposal functions. We therefore propose to use recurrence intervals for comparing the mixing properties of different proposal functions. Define two disjoint subsets $A$ and $B$ of the state space that are far apart from each other and estimate the expected number of iterations needed to move between these two subsets. The Markov chain is split into recurrence intervals characterized by when the chain enters into $A$, then enters into $B$ and the interval is ended when the chain again returns to $A$. More formally, we define the indices $i_1 < i_2 < i_3 < \cdots$ where $i_k$ is the first time the chain enters into $A$ after the chain has been in $B$ and the subindex $k$ is used the k'th time this happens. Define the non-overlapping intervals, $I_1, I_2, I_3, \ldots$ where $I_k$ is the interval $(i_k, i_{k+1} - 1)$. The intervals $I_k$ satisfy the following properties: (i) $x_{i_k} \in A$ (ii) there is at least on state $x_i \in B$ with $i_k < i < i_{k+1}$ (iii) for all $x_i \in B$ we have $x_j \notin A$ for $i_k < i < j < i_{k+1}$ and (iv) $x_{i_{k+1}} \in A$. From the definition it is easy to see that the length of all the recurrence intervals are from the same distribution. We define $L_k = i_{k+1} - i_k$, the length of k'th interval $I_k$ and $M_{A,B} = E\ L_k$, the expected length of the interval.

It is most interesting to choose $A$ and $B$ far apart from each other where we expect it is difficult to move between the two subsets. If we don't know about particular problems in the mixing, we may choose two arbitrary subsets $A$ and $B$ that far apart from each other. For example, we may select the first component $x_{,1}$, and choose $A = \{x|\ x_{,1} < a\}$ and $B = \{x|\ x_{,1} > b\}$ for $a < b$. Alternatively, we may include all components in the definition of $A$ and $B$ choosing $A = \{x|\ x_{,i} < a_i\ for\ all\ i\ \}$ and $B = \{x|\ x_{,i} > b_i\ for\ all\ i\}$. It is easier to argue for negligible correlation between parameters from different recurrence intervals if all components have a large change in value. We estimate $\widehat{M_{A,B}} = n/m_{A,B,n}$ from a (long) Markov chain of length $n$ where $m_{A,B,n} = k$ is the number of times the chain has entered $A$ then moved to $B$ then returned to $A$. Using the indexes in the definition of the recurrence intervals, we have $i_k$ is the largest index such that $i_k \leq m_{A,B,n}$. This may also be described as the longest sequence of indices $i_1 < j_1 < i_2 < j_2 \ldots < i_k$ that satisfy $x_{i_1} \in A$, $x_{j_1} \in B$, $x_{i_2} \in A$, $x_{j_2} \in B, \ldots, x_{i_k} \in A$. In this paper we have chosen $A$ and $B$ in opposite parts of the state space and with $\pi(A) \approx \pi(B)$ making it easier to argue that recurrence periods are close to independent of each other. We could also choose $A$ far from the center of the limiting density with $\pi(A)$ small, $A \cap B = \emptyset$ and $B$ such that $\pi(B) = 0.5$.

Below we show some properties of $M_{A,B}$. It is well suited to identify the areas that it is difficult to sample from.

### Proposition

If the proposal function is $\pi(A)$ in each iteration, then $M_{A,B} = \frac{1}{\pi(A)} + \frac{1}{\pi(B)}$.

The proof is given in the Appendix.

Notice that it is possible to make $M_{A,B}$ arbitrary large by choosing $\pi(A)$ or $\pi(B)$ sufficient small. We are interested in the situation where it is difficult to move between $A$ and $B$ but without $\pi(A)$ and $\pi(B)$ being too small making it necessary with a very long Markov chain to estimate $M_{A,B}$. If there are subsets $A$ and $B$ where we are particular interested in knowing $\pi(A)$ and $\pi(B)$, this may be a good choice since the Theorem in the next Section gives a bound on the variance of the estimate of $\pi(A)$ and $\pi(B)$. We expect $H_{A,B} = \frac{M_{A,B}}{\frac{1}{\pi(A)} + \frac{1}{\pi(B)}} > 1$ since other proposal functions does not sample the state space as efficient as the limiting density. However, it is possible to find artificial examples where $H_{A,B} < 1$. See appendix. If we find a pair $A$, $B$ where $H_{A,B} \gg 1$, we may conclude that the mixing is poor. It is easy to estimate $H_{A,B}$ for a given $A$ and $B$ from a long Markov chain, but to find $A$ and $B$ that make $H_{A,B}$ large may be more challenging.

## 3. Probability estimate

The obvious probability estimated from a Markov chain is $\widehat{\pi_n(A)} = \frac{1}{n}\{\# \ x_i \in A, for \ 0 < i \leq n\}$, the number of states in $A$ after $n$ iterations divided by $n$. It is well-known that $E\widehat{\pi_n(A)} = \pi(A)$, assuming we have already reached convergence. We will estimate the variance of $\widehat{\pi_n(A)}$ based on $m_{A,B,n}$, defined above as the number of times a Markov chain has moved between $A$ and $B$ in $n$ iterations. We will show that this is a good measure on how good the $n$ elements of the Markov chain represent the limiting distribution. We will use the non-overlapping recurrence intervals, $I_1, I_2, I_3, \ldots$ where $I_k$ has the indices $(i_k, i_{k+1} - 1)$ defined above. Further, define $P_k = \{\# \ x_i \in I_k \cap A\}$, the number of states in $A$ in the period $I_k$, and $R_k = \frac{P_k}{L_k \pi(A)}$ such that we expect $R_k \approx 1$. This is the fraction of states in $I_k$ that is inside $A$ divided by $\pi(A)$. For $n = i_k$, i.e. after exactly $k$ intervals, then $\widehat{\pi_n(A)} = \pi(A) \sum_{j=1}^{k} R_j \frac{L_j}{\sum_{s=1}^{k} L_s}$. Notice that $\pi(A)$ is in the nominator of $R_s$ making the expression independent of $\pi(A)$. We may formulate the following Theorem.

## Theorem

Assume that:

(i) there is a $j < 0$ such that $x_j \in B$ is from the target density $\pi(x)$ restricted to subset $B$;

(ii) $x_s \notin A$ for $j < s < 0$ and $x_0 \in A$ and

(iii) $COV\left((\frac{1}{\sum_{s=1}^k L_s})^2, \left(\sum_{j=1}^k (P_j - \pi(A)L_j)\right)^2\right) < 0$,

then for $n = i_k$

$$VAR(\widehat{\pi_n(A)}) \leq \pi^2(A) \, E(\frac{k \, M_{A,B}}{\sum_{s=1}^k L_s})^2 VAR\left(\sum_{j=1}^k (R_j - 1)\frac{L_j}{kM_{A,B}}\right).$$

If in addition the covariance $COV\left(\frac{P_i}{\pi(A)} - L_i, \frac{P_j}{\pi(A)} - L_j\right) \leq c2^{-|i-j|-1} VAR\left(\frac{P_1}{\pi(A)} - L_1\right)$ for a constant $c \geq 0$ and all indices $i \neq j$, then

$$VAR(\widehat{\pi_n(A)}) \leq \pi^2(A) \, \frac{1+c}{k} E\left(\frac{k \, M_{A,B}}{\sum_{s=1}^k L_s}\right)^2 VAR\left((R_1 - 1)\frac{L_1}{M_{A,B}}\right).$$

The proof is given in the Appendix. The Theorem is a generalization of the trivial result

$$VAR\left(\frac{1}{k}\sum_{i=1}^k X_i\right) = \frac{(EX_1)^2}{k} VAR\left(\frac{X_1}{EX_1}\right)$$

for i.i.d. variables $X_i$. We have divided by $EX_1$ in the variance in order to show the dependency on $EX_1$ when this vanish. Here $X_i$ is our estimate on $\pi(A)$ based on the interval $I_i$.

We have the same bound for $VAR(\widehat{\pi_n(B)})$. The two first requirements is satisfied if $x_j$ is from the target density $\pi(x)$, then the chain enters into $B$, and $x_0$ is the first state in the Markov chain in $A$ after it has been in $B$. This is the same as $i_1 = 0$ according to the notation in the previous section. This assumption is made in order to have the same statistical properties for all the intervals $I_k$.

The assumption that $COV\left((\frac{1}{\sum_{s=1}^k L_s})^2, \left(\sum_{j=1}^k (P_j - \pi(A)L_j)\right)^2\right) < 0$ is reasonable, particularly if the subsets $A$ and $B$ are far from each other in the state space. We have $E(P_j - \pi(A)L_j) = 0$. As we illustrate later, the distribution of $L_j$ has a heavy tail. The heavy tail dominates the distribution of $(P_j - \pi(A)L_j)^2$. Then we have $COV(L_j^2, (P_j - \pi(A)L_j)^2) > 0$ and $COV\left(\frac{1}{L_j^2}, (P_j - \pi(A)L_j)^2\right) < 0$. The intervals $I_1, I_2, I_3, \ldots$ are almost independent making $COV\left((\frac{1}{\sum_{s=1}^k L_s})^2, \left(\sum_{j=1}^k (P_j - \pi(A)L_j)\right)^2\right)$ small and negative. If $COV\left((\frac{1}{\sum_{s=1}^k L_s})^2, \left(\sum_{j=1}^k (P_j - \pi(A)L_j)\right)^2\right) = 0$ then

$$VAR(\widehat{\pi_n(A)}) = \frac{\pi^2(A)}{k^2} E(\frac{k \, M_{A,B}}{\sum_{s=1}^k L_s})^2 VAR\left(\sum_{j=1}^k (R_j - 1)\frac{L_j}{M_{A,B}}\right).$$

If the covariance expression had been positive, this would only make the upper bound in the Theorem slightly larger. It is reasonable that the covariance $COV\left((R_i - 1)L_i, (R_j - 1)L_j\right) \approx 0$ and decreasing exponentially with $-|i-j|$ since the intervals are almost independent and the correlation in a Markov chain decreases exponentially with the distance. This gives the second bound in the theorem.

We expect $E(\frac{k\, M_{A,B}}{\sum_{s=1}^{k} L_s})^2$ is close to 1 and $VAR\left((R_1 - 1)\frac{L_1}{M_{A,B}}\right)$ is reasonable small, implying that we expect $VAR(\widehat{\pi_n(A)})$ to be at the order $\pi^2(A)/k$. For the multi-normal model in Example 1 and the climate model in Example 4, we have $VAR(\widehat{\pi_n(A)}) < 4\pi^2(A)/k$, see Figure 3 and 5. The Theorem shows that $VAR(\widehat{\pi(A)})$ is proportional with $M_{A,B}$ and $H_{A,B}$ since $k \approx n/M_{A,B}$. This shows the importance of adjusting the simulation parameters in order to minimize $M_{A,B}$ and $H_{A,B}$.

## 4. Acceptance rates

Roberts et al. (1997) proved the remarkable result that if the target density is on the form

$$\pi(x_{.,1}, x_{.,2}, \ldots, x_{.,d}) = f(x_{.,1})f(x_{.,2})\ldots f(x_{.,d})$$

then as $d \to \infty$ the optimal acceptance rate is 0.234 and in fact the optimal acceptance rate is close to 0.234 already for $d > 5$. Here $x \in \mathcal{R}^d$ and we use the notation $x = (x_{.,1}, x_{.,2}, \ldots, x_{.,d})$ to describe the different components. For $d = 1$, the optimal acceptance rate for the normal distribution is 0.44. Numerical studies (e.g. Gelman et al (1996)) show that the algorithm is reasonable efficient for acceptance rate in the range (0.1,0.6). Roberts and Rosenthal (2001) generalize the result also to inhomogeneous target densities on the form

$$\pi(x_{.,1}, x_{.,2}, \ldots, x_{.,d}) = \prod_{i=1}^{d} C_i f(C_i x_{.,i})$$

under some additional conditions. See also Neal and Roberts (2006) for generalization regarding the dimensionality of the updating rule. Rosenthal (2010) gives the general advice to target an acceptance rate in the range specified above. This advice also includes multi-modal target densities.

We give examples where the optimal acceptance rate is much smaller and in fact vanishes in the limits. Instead of the very general advice to have the acceptance rate in the range (0.1,0.6), our advice is to

*use simulation parameters, typically the step length in the proposal function, that make $H_{A,B}$ small for fixed A and B.*

It is equivalent to make $H_{A,B}$ and $M_{A,B}$ small for fixed $A$ and $B$. $H_{A,B}$ depends on $A$ and $B$ and it is not critical to minimize $H_{A,B}$, only to find values that make $H_{A,B}$ reasonable close to the minimum. Our experience, however from a limited number of models, is that the simulation parameters that minimize $H_{A,B}$, fortunately do not critically depend on the choice of $A$ and $B$. See Figure 1 as an example. This advice is closely connected to increase the mixing of the chain and to reduce $VAR(\widehat{\pi_n(A)})$.

### Example 1, the multi-normal case

Here the state space is in d-dimensions and the target density is the product of $d$ normal densities $((x_{.,1}, x_{.,2}, \ldots, x_{.,d})) = (\varphi(x_{.,1}/\sigma_1)/\sigma_1) \prod_{i=2}^{d} \varphi(x_{.,i})$. When $\sigma_1$ is small, the target density has very different scale in the first dimension compared to the other dimensions. We assume this difference in scale is not known and the proposal function is

$$q\big((y_{.,1}, y_{.,2}, \ldots, y_{.,d}), (x_{.,1}, x_{.,2}, \ldots, x_{.,d})\big) = \prod_{i=1}^{d} (\varphi((y_{.,i} - x_{.,i})/\sigma_2)/\sigma_2)$$

If we chose $\sigma_1$ small, then the scale is very different in the different directions of the state space making it difficult for the Markov chain to converge to the limiting distribution. Figure 2, left panel shows $H_{A,B}$ when varying the boundaries $A$ and $B$ and $\sigma_2$. Notice that the same value of $\sigma_2$ minimize $H_{A,B}$ for all values of $A$ and $B$ and that $H_{A,B}$ is decreasing when the subset $A$ and $B$ becomes more extreme. Right panel shows how $M_{A,B}$ depends on $\sigma_1$ and $\sigma_2$.

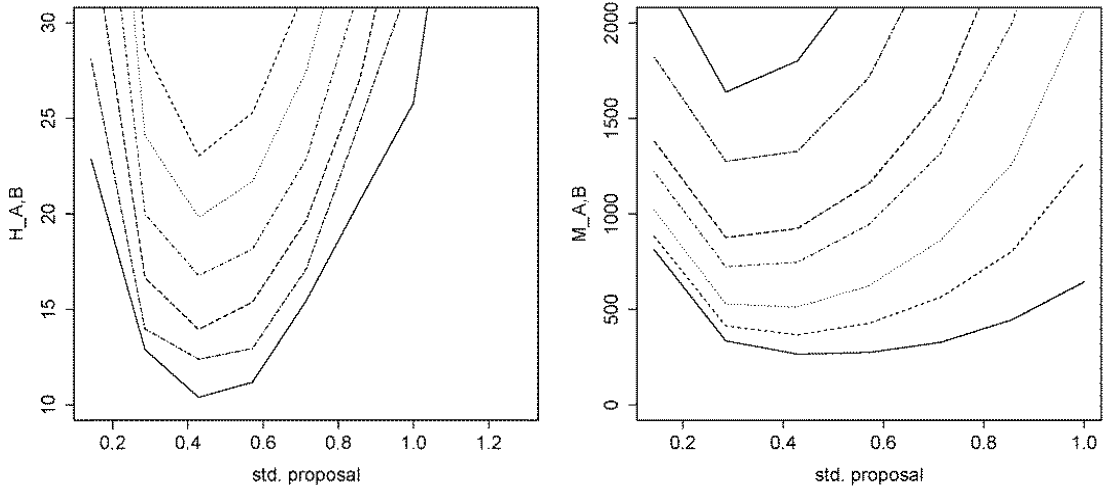

Figure 1. Example 1 with multi-normal limiting density. Left panel shows $H_{A,B}$ when varying $\sigma_2$ and $A_i$ and $B_i$. Each curve is for different subset areas $A_i$ and $B_i$ with threshold for 0.75, 1, 1.25, 1.5, 1.75, 2, 2.25. $H_{A,B}$ is decreasing for more extreme threshold. Right panel shows $M_{A,B}$ when varying $\sigma_2$ and $\sigma_1$. Each curve is for $\sigma_1 = 1, \frac{1}{2}, \frac{1}{3}, \frac{1}{5}, \frac{1}{8}, \frac{1}{10}, \frac{1}{15}, \frac{1}{20}, \frac{1}{50}$ in the limiting density. $M_{A,B}$ is increasing when $\sigma_1$ decreases. Curves are based on 300.000 iterations in the Markov chain.

## 5. Analyzing the chain between A and $B$

This section discusses the challenges for the Markov chain to move between the areas $A$ and $B$. We have shown that a large $H_{A,B}$ value indicates poor mixing and makes it necessary with a large number of iterations in order to reduce $VAR\big(\widehat{\pi_n(A)}\big)$.

We need some notation in order to study the case when there is poor mixing. Given a state $x_0 \in A$, it is possible to define a sequence of subsets $C = \{C_1, C_2, \ldots, C_{n-1}\}$ such that the probability for a Markov chain starting in $x_0 \in A$ with $x_i \in C_i$ and $x_n \in B$ is not too low. In order to make this a likely Markov chain $C$ must depend on the proposal function. If the movement from $A$ to $B$ requires a gradual change along the $x_{.,1}$-axis (see Example 3 later), we may choose $C_i = \{x | x_{1,a} + (i-1)\sigma_2 < x_{.,1} < x_{1,a} + i\sigma_2\}$. This requires an average increase with $\sigma_2$ in the first component in each iteration of the Markov chain. If $H_{A,B} \gg 1$, then either the chain $C$ is very long ($n$ large) and/or some of the steps have very small probability. We discuss each of these cases separately.

If there are several modes in the target density, then usually the challenge is to move between the modes. This may require that some states in the Markov chain have very small target density or it is necessary with long jumps in order to avoid these areas. A combination of the two alternatives is also possible. This is illustrated in Example 2 below where the limiting density has two modes. In this example it is optimal to choose long jumps making the acceptance rate very small when the distance between the modes is large.

Example 3 below with difference in scale illustrates the situation where it is necessary with very small acceptance rate and many steps in order to move between the two subsets in the state space. However, for a huge state space with many variables it may be reasonable with large values of $H_{A,B}$ and this is not necessarily a sign on poor mixing. As an illustration, the distance in the unit-box in $R^d$ between (0,0,…,0) and (1,1,…,1) is $d^{0.5}$ while the distance between (0,0,0,…,0) and (1,0,0,…,0) is 1. This is a major difference for $d$ large. When we judge whether a Markov chain has poor mixing, it is not sufficient to only consider the value of $H_{A,B}$, we must also take into consideration the size of the state space.

Below we show two toy examples with a continuous target distribution where it is optimal to have very small acceptance rate. Then it is given a large climate model documented in other papers illustrating the use of the technique presented in this paper. We also compare $S_1 = E|x_{i+s} - x_i|$ for the different models.

### Example 2. Two modes:

This example is in one dimension and with target density $\pi(x) = (\varphi(x) + \varphi(x + a))/2$ where $\varphi(x)$ is the normal density N(0,1). This target density has two modes $x = 0$ and $x = a$ and it is increasingly difficult for the Markov chain to move between the two modes for increasing values of $a$. Let the proposal function be $q(y, x) = \varphi((x - y)/\sigma)/\sigma$. If $a$ is large compared to $\sigma$, the Metropolis-Hastings algorithm uses many iterations in order to move between the two modes. The properties of this model are shown in Table 1. Each line shows the result for $\sigma$ that minimizes $M_{A,B}$ for the specified target distribution (i.e. the chosen $a$). A large value of $\sigma$ reduces the acceptance rate but increases the probability for a move between the two modes. Hence, it is possible to find arbitrary small optimal acceptance rates by setting the constant $a$ large enough. For one dimension minimizing $M_{A,B}$ is almost the same as optimizing $S_1$. However, if we generalize to d-dimensions then there may be a major difference. We may obtain the optimal $S_1$ within one of the modes with steps such that the problem of moving to the other mode is minimal.

| $a$ | Optimal $\sigma$ | Acceptance rate | $S_1$ | $M_{A,B}$ | $H_{A,B}$ |
|---|---|---|---|---|---|
| 2 | 3.25 | 0.62 | 1.01 | 9.0 | 2.3 |
| 4 | 5.5 | 0.35 | 1.51 | 16.8 | 3.9 |
| 6 | 7.5 | 0.24 | 1.84 | 24.6 | 5.7 |
| 8 | 9.5 | 0.18 | 2.12 | 32.8 | 7.5 |
| 10 | 12.3 | 0.14 | 2.36 | 40.4 | 9.3 |
| 12 | 14.3 | 0.12 | 2.60 | 47.8 | 11 |
| 14 | 14.0 | 0.11 | 2.80 | 56.0 | 13 |

Table 1. Example 2 with two modes. The Table shows for each value $a$, the value of $\sigma$ that minimize $M_{A,B}$ and the corresponding acceptance rate and mean jump i.e. $S_1$. Here $A = \{x| x < 1\}$ and $B = \{x| x > a - 1\}$. Data is based on 100.000 simulations, but this may not be large enough that the last digit is correct.

## Example 3. Problem of scale

Here the state space is in two dimensions and the target density $\pi((x_{,1},x_{,2})) = \varphi(x_{,1})\varphi(x_{,2}/\sigma_1)/\sigma_1$ for $x_{,1} > x_{,2}$ and $\pi((x_{,1},x_{,2})) = \varphi(x_{,1}/\sigma_1)\varphi(x_{,2})/\sigma_1$ otherwise. When $\sigma_1$ is small, the target density has very different scale in the two dimensions. The target density is continuous and varies fast with $x_{,2}$ for $x_{,1} > x_{,2}$ and varies fast with $x_{,1}$ otherwise. Since the scale varies in the state space, it is not easy to handle this in a random walk proposal function. Let the proposal function be $q((y_{,1},y_{,2})(x_{,1},x_{,2})) = \varphi((x_{,1} - y_{,1})/\sigma_2)\varphi((x_{,2} - y_{,2})/\sigma_2)/\sigma_2^2$. If we chose $\sigma_1$ small, then the scale is very different in different parts of the state space making it difficult for the Markov chain to converge to the limiting distribution. The properties of this model are shown in Table 2. Each line shows the result for $\sigma_2$ that minimizes $M_{A,B}$ for the specified target distribution (i.e. the chosen $\sigma_1$). There is a difficult trade off when setting the standard deviation $\sigma_2$ in the proposal function. It is necessary to have it quite small in order to get a satisfactory acceptance rate, but then the Markov chain moves very slowly in the direction where the limiting function varies slowly. Also here we may find arbitrary small optimal acceptance rates by setting the $\sigma_1$ small enough.

| $\sigma_1$ | Optimal $\sigma_2$ | Acceptance rate | $S_1$ | $M_{A,B}$ | $H_{A,B}$ |
|---|---|---|---|---|---|
| 1 | 1.5 | 0.47 | 0.81 | 20 | 3.6 |
| 0.5 | 1.0 | 0.46 | 0.51 | 24 | 3.7 |
| 0.25 | 1.0 | 0.36 | 0.42 | 32 | 3.7 |
| 0.1 | 0.9 | 0.24 | 0.30 | 66 | 5.6 |
| 0.05 | 0.9 | 0.21 | 0.20 | 120 | 8.8 |
| 0.01 | 0.6 | 0.056 | 0.089 | 560 | 41 |
| 0.002 | 0.5 | 0.011 | 0.038 | 2 700 | 210 |
| 0.001 | 0.6 | 0.0042 | 0.028 | 4 700 | 300 |

Table 2. Example 3 with problem with scale. The Table shows for each value $\sigma_1$, the value of $\sigma_2$ that minimize $M_{A,B}$ and the corresponding acceptance rate and mean jump i.e. $S_1$. Here $A = \{x | x_{,1} < -0.4\}$ and $B = \{x | x_{,1} > 0.4\}$. Data is based on 100.000 simulations, but this may not be large enough that the last digit is correct.

## Example 4. The climate model

The climate model is a complex non-linear model that is sampled by a Markov chain and documented in Aldrin et al. 2012 and Skeie et al. 2014. The model has a large number of parameters and we know that the mixing of the most important response parameter, the climate sensitivity, is slow. Figure 2, left panel shows the slow mixing of the climate sensitivity in 100.000 elements of the Markov chain. This makes it necessary to run the model for weeks in order to get a good estimate on the distribution of the climate sensitivity which is the main objective of the model. It is not possible to rewrite the sampling to an adaptive model. The Markov chain is a Metropolis-Hastings sampler where the parameters are updated in blocks. The parameter blocks are updated either by a random-walk update (25 parameter blocks) or by using a Gibbs sampler. A random-walk update is used when the prior distribution is normal or uniform, while a Gibbs-sampler update is used when the prior is gamma or Wishart. The step length in the random walk updates are adjusted in the burn in period in order to get a 0.3 acceptance rate in the previous published papers on the model.

Figure 2 right panel shows $H_{A,B}$ for different pairs of thresholds and different acceptance rates for group of parameters with the climate sensitivity parameter. The other parameters groups may have other acceptance rates but these parameters are not as important for the mixing of the Markov chain. The figure indicates that we obtain the smallest $H_{A,B}$ values for acceptance rates in the interval

(0.1,0.35) and that $H_{A,B}$ increases slightly for more extreme values of $A_i$ and $B_i$. In this paper we have tested other acceptance rates and found out that 0.07 acceptance rate in all blocks except the block with the climate sensitivity and 0.14 acceptance rate in this block is more efficient. Table 3 shows that results with this acceptance rate.

These examples show the importance of scaling the proposal function such that $M_{A,B}$ is as small as possible.

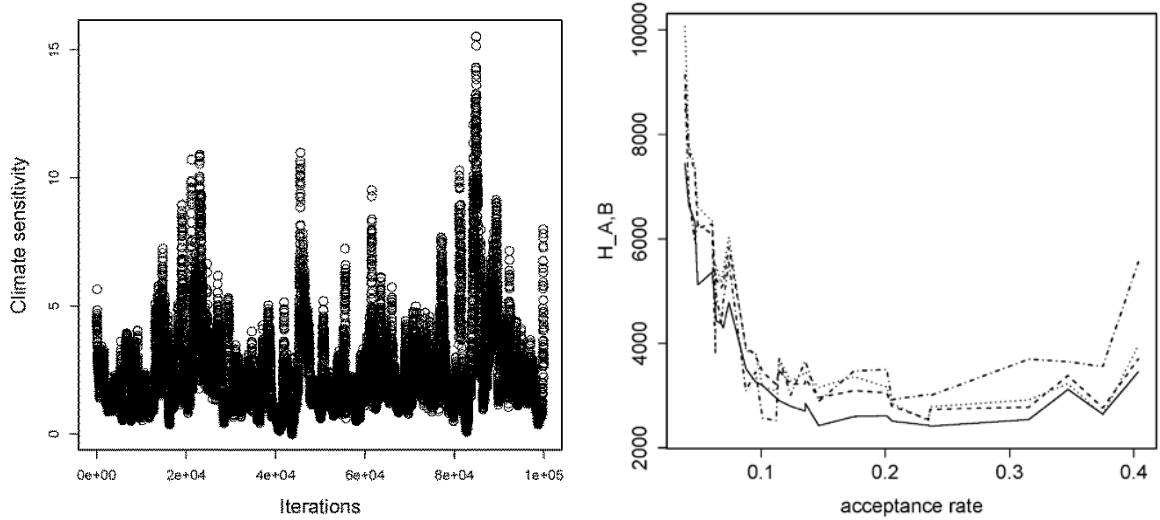

Figure 2 The climate sensitivity in the climate model. Left: the climate sensitivity in 100.000 iterations of the Markov chain showing slow mixing. Only each 50$^{th}$ element in the chain is shown in the plot. Right: The figure shows $H_{A,B}$ when the model is tuned to different acceptance rates for the climate sensitivity parameter. Each curve is for different pairs of $A_i$ and $B_i$ and at the horizontal axis is the value of the acceptance rate for the parameter group with the climate sensitivity. The curves are more dotted for more extreme values of $A_i$ and $B_i$. Estimates are based on 28 runs each with more than one month CPU time and about 75 mill. iterations after burn in.

| $\pi(A_i)$ | $\pi(B_i)$ | $M_{A,B}$ | $H_{A,B}$ |
|---|---|---|---|
| 0.01 | 0.03 | 275 000 | 2 400 |
| 0.05 | 0.10 | 86 000 | 2 700 |
| 0.11 | 0.20 | 42 000 | 3 000 |
| 0.16 | 0.28 | 29 000 | 3 000 |
| 0.23 | 0.39 | 16 600 | 2 400 |

Table 3. The results from five different thresholds of the climate sensitivity in the climate model. Notice the large number of iterations in the Markov chain in order to move from the extreme values of the climate sensitivity. This is based on 75 mill. iterations.

## 6. Statistical properties of the chain between A and B

In section 2 we defined the non-overlapping recurrent intervals $I_1, I_2, I_3, \ldots$ of the Markov chain where each period starts when the chain enters into $A$, then enters into $B$ and the period is ended when the chain returns to $A$. All the periods have the same statistical properties and we have defined $M_{A,B} = E\, L_k$ the expected length of a period and $R_k$ as the fraction of states in $I_k$ that is inside $A$ divided by $\pi(A)$. The statistical properties of the intervals are important for $VAR(\widehat{\pi_n(A)})$ according to the Theorem. We have not succeeded in proving general properties of the stochastic variables $L_k$ and $R_k$. We limit our self to empirical studies on a multi-normal model and the climate model and two smaller more "extreme" cases. However, we believe these properties are quite general.

In our two examples, $R_k$ is close to an exponential distribution with parameter 1. Our experience is that the density of the length of the intervals $L_k$ is heavily tailed that we prefer to model with the Weibull distribution. In our two examples and the first extreme case we used a Weibull parameter $k \approx 1.5$ which gives a distribution that is less heavy tailed than the exponential distribution. In the last extreme case given at the end of the section, the distribution is more heavy tailed then the exponential distribution.

### Example 1, the multi-normal case, continued

The properties of this model are shown in Figure 2 and Table 3. Each line shows the result for the value of $\sigma_2$ that minimizes $M_{A,B}$ for the specified target distribution (i.e. the chosen $\sigma_1$). There is a trade off when setting the standard deviation $\sigma_2$ in the proposal function. It is necessary to have it quite small in order to get a satisfactory acceptance rate, but then the Markov chain moves very slowly in the direction where the limiting function varies slowly. Also here we may find arbitrary small optimal acceptance rates by setting the $\sigma_1$ small enough. Notice that the optimal value of $\sigma_2$ decreases slightly when $\sigma_1$ decreases but the decrease is not large enough such that optimal acceptance rates decreases.

| D | $\sigma_1$ | Optimal $\sigma_2$ | Acceptance rate | A | P(A) | $M_{A,B}$ | $H_{A,B}$ |
|---|---|---|---|---|---|---|---|
| 3 | 1 | 0.87 | 0.36 | 2 | 0.0024 | 2760 | 3.0 |
| 3 | 0.33 | 0.67 | 0.27 | 2 | 0.0024 | 4230 | 4.6 |
| 3 | 0.2 | 0.61 | 0.20 | 2 | 0.0024 | 6500 | 7.2 |
| 3 | 0.125 | 0.56 | 0.15 | 2 | 0.0024 | 4000 | 9.4 |
| 3 | 0.1 | 0.55 | 0.12 | 2 | 0.0024 | 3700 | 12 |
| 3 | 1 | 0.87 | 0.36 | 1 | 0.079 | 110 | 4.5 |
| 3 | 0.33 | 0.67 | 0.27 | 1 | 0.079 | 205 | 8.1 |
| 3 | 0.2 | 0.61 | 0.20 | 1 | 0.079 | 305 | 12 |
| 3 | 0.125 | 0.56 | 0.15 | 1 | 0.079 | 460 | 18 |
| 3 | 0.1 | 0.55 | 0.12 | 1 | 0.079 | 570 | 23 |
| 10 | 1 | 0.48 | 0.31 | 1 | 0.079 | 270 | 11 |
| 10 | 0.33 | 0.42 | 0.27 | 1 | 0.079 | 370 | 15 |
| 10 | 0.2 | 0.38 | 0.24 | 1 | 0.079 | 520 | 20 |
| 10 | 0.125 | 0.35 | 0.19 | 1 | 0.079 | 730 | 29 |
| 10 | 0.1 | 0.35 | 0.16 | 1 | 0.079 | 900 | 35 |
| 10 | 0.066 | 0.34 | 0.11 | 1 | 0.079 | 1 300 | 51 |
| 10 | 0.05 | 0.33 | 0.091 | 1 | 0.079 | 1 700 | 67 |
| 10 | 0.02 | 0.34 | 0.035 | 1 | 0.079 | 4 300 | 170 |
| 10 | 0.1 | 0.35 | 0.16 | 1.5 | 0.016 | 2 700 | 24 |
| 10 | 0.1 | 0.35 | 0.16 | 2 | 0.0024 | 13 700 | 16 |

Table 3. Example 1 the multi-normal case and shows the optimal $\sigma_2$ that minimizes $M_{A,B}$. Here $A = \{x| \, x_{.,1} < -a\}$ and $B = \{x| \, x_{.,1} > a\}$. The simulation is based on one chain with length 300.000. This is too little to get a good estimate on the optimal $\sigma_2$ and to estimate $M_{A,B}$ when it is large.

Note that $H_{A,B}$ decreases when we make the two subset $A$ and $B$ more extreme by increasing $a$ as shown in Figure 1, left panel. This is our general experience. Hence, the tail of the limiting distribution does not seem to be critical provided the proposal function is scaled properly.

We have estimated $Var(\widehat{\frac{\pi_n(A)}{\pi(A)}})$, see Figure 3. We have simulated 1.000 chains in $100M_{A,B}$ iterations and estimated the decrease in the standard deviation as the number of iterations increases. Notice the similarity of the curves for a wide range of $M_{A,B}$ from 264 to 4760. We obtain smaller standard deviations for increasing $M_{A,B}$ which comes from smaller values of $\sigma_1$ and larger values of $a$. Figure 3, right panel, shows the distribution of $\widehat{\frac{\pi_n(A)}{\pi(A)}}$ for $n = 100M_{A,B}$. Since this estimate is the weighted average of 100 $R_k$ variables, it is close to a normal density.

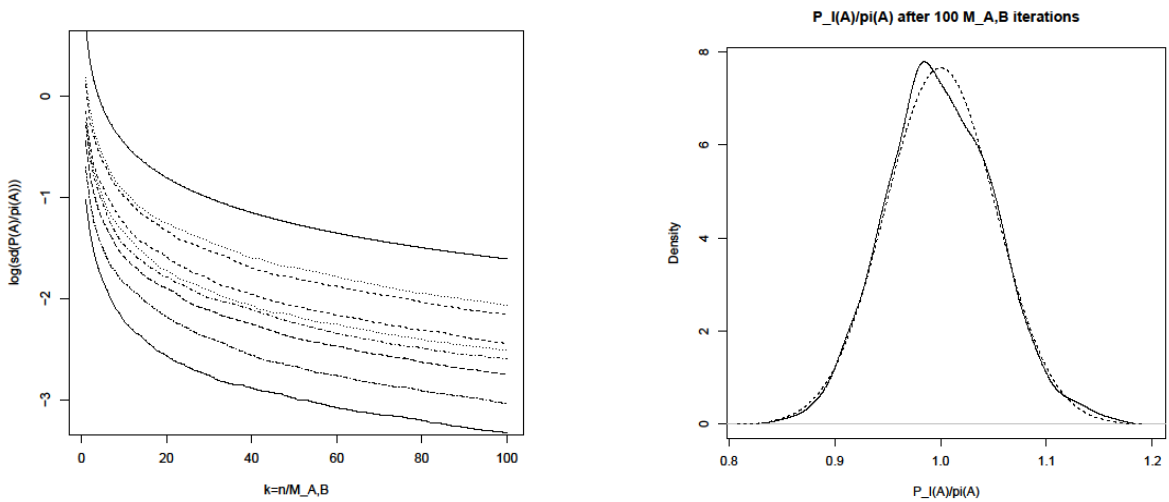

Figure 3. The left panel is the logarithm of the standard deviation of $\widehat{\frac{\pi_n(A)}{\pi(A)}}$ in the multi-normal model with $d = 10$ for $k = \frac{n}{M_{A,B}} = 1,2,\dots,100$ with $\sigma_1 = 1, \frac{1}{5}, \frac{1}{8}, \frac{1}{10}, \frac{1}{20}, \frac{1}{50}$ and $a = 1$ and with $\sigma_1 = \frac{1}{10}$ for both $a = 1.5$ and $a = 2$. The values are decreasing with decreasing values of $\sigma_1$. The upper curve is the $\log(\frac{2}{\sqrt{k}})$ function. Estimates are based on 1.000 chains. The right panel is the distribution of $\widehat{\frac{\pi_n(A)}{\pi(A)}}$ for $\sigma_1 = \frac{1}{50}$ and $a = 1$ after $100M_{A,B} = 4.300.000$ iterations where $\pi(A) = 0.079$ and the normal fit to the density.

We have found the distribution of the length of the intervals $L_A$ and $R_A$, see Figure 4. This is the same example as line 13 in Table 3. Both these distributions are satisfactory fitted with a Weibull distribution. The cross plot at the bottom of Figure 5 shows a negative correlation between $L_k$ and $P_k - \pi(A)L_k$. Here $cor(L_k, P_k - \pi(A)L_k) = -0.65$ and $cor(L_k^{-2}, (P_k - \pi(A)L_k)^2) = -0.10$ showing that the assumptions in the Theorem is satisfied in this case.

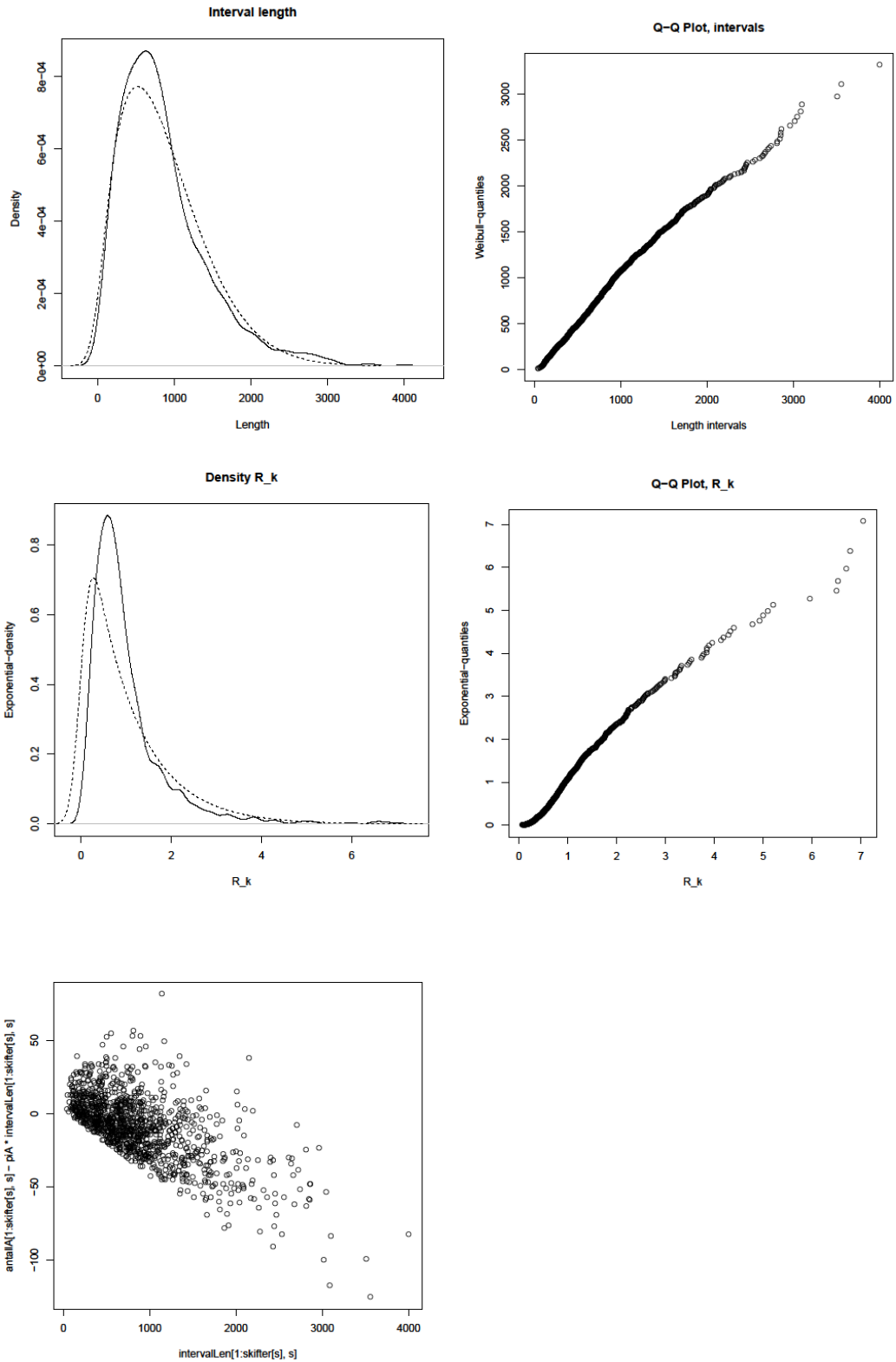

Figure 4. From the multi-normal model $d = 10$ and 1 mill. iterations. $\pi(A) = 0.07$, and $M_{A,B} = 474$. First line is the density and a qq-plot of the length $L_k$ fitted to a Weibull distribution with parameters 1.57 and 922. Second line is the fraction of $R_k$ fitted to an exponential distribution with parameter 1 since $ER_k \approx 1$. Lower panel is a cross-plot of the $L_k$ and $P_k - \pi(A)L_k$. This is based on 1171 intervals.

## Example 4. The climate model, continued

We have estimated $Var(\frac{\widehat{\pi_n(A)}}{\pi(A)})$ in the climate model, see Figure 5. The figure shows that the standard deviation of the estimate decreases as described in the Theorem.

Also in the climate model the length of the intervals $L_k$ is a heavily tailed density that is satisfactory model with the Weibull distribution. The ratio $R_k$ is best fitted with an exponential distribution. The heavier the tail of $L_k$ is, the longer the Markov chain must be in order to represent the target density. Figure 6 shows an example with the density of $L_k$, density of $R_k$ and a cross-plot of the $L_k$ and $P_k - \pi(A)L_k$ based on more than 1 953 intervals. We have tested more than 130 combinations of parameters and subset $A_i$ and $B_i$ and estimated $cor(L_k^{-2}, (P_k - \pi(A)L_k)^2) < 0$ except in one case where an outlier clearly dominated the estimate.

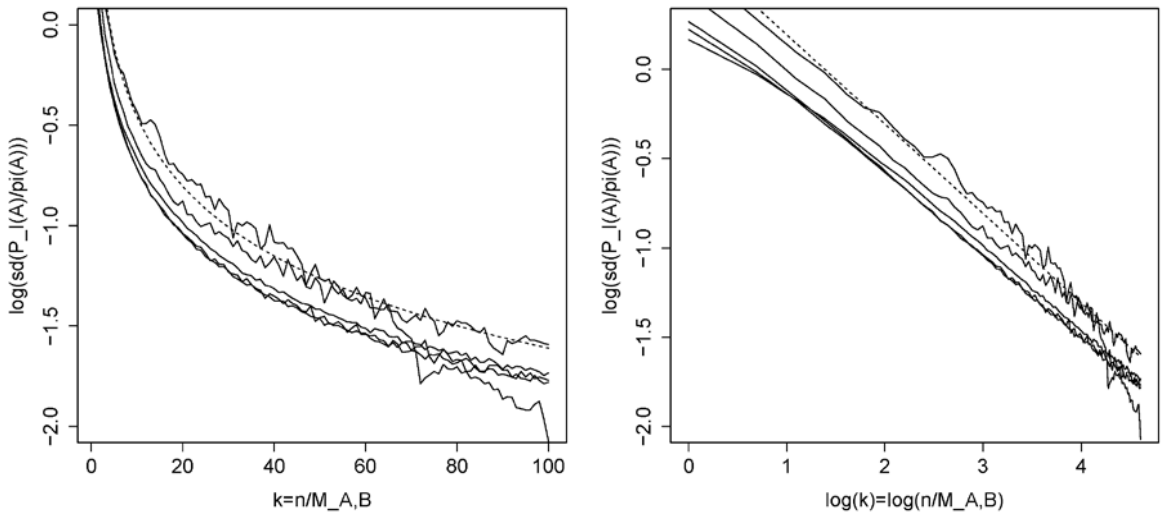

Figure 5. Left panel: The logarithm of the standard deviation of $\frac{\widehat{\pi_n(A)}}{\pi(A)}$ in the climate model for $k = \frac{n}{M_{A,B}} = 1,2,…,100$. There is one curve for each of the 5 $A_i, B_i$ pairs shown in Table 4. The curve is estimated from 28 long chains each simulated in about 1 month CPU time. These chains that are cut into sections of length $k = n/M_{A,B}$ The estimate is based on fewer sections for larger values of k and the variability is larger for the most extreme $A_i, B_i$ pair where we have less data. The dashed line is the $\log(\frac{2}{\sqrt{k}})$ function. The right panel is a log-log plot of the same figure. Estimates are based on 28 runs each with more than one month CPU time and about 75 mill. iterations after burn in.

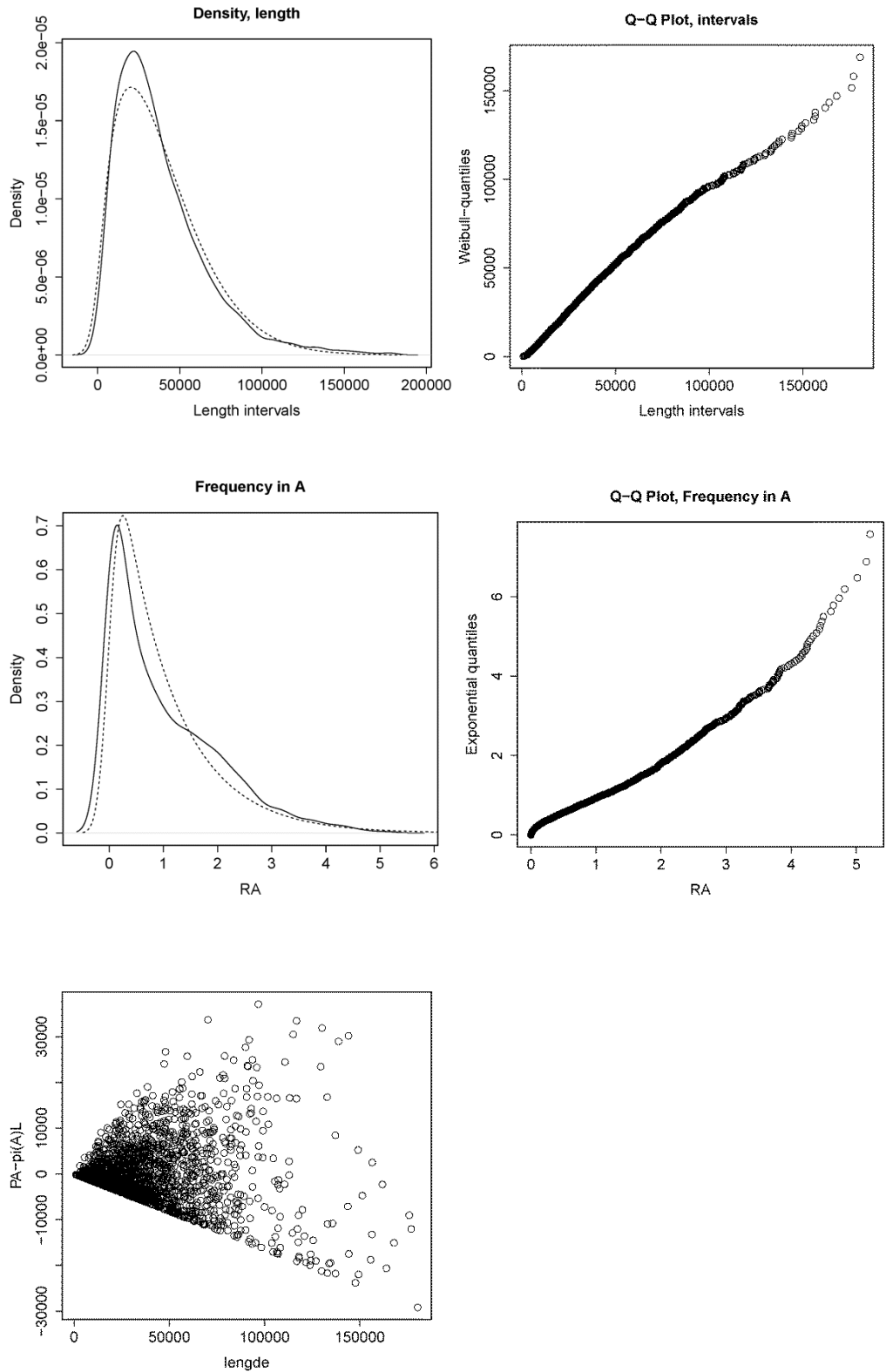

Figure 6: Similar to Figure 4, but for climate model. First line is the density and a qq-plot of the length $L_k$ fitted to a Weibull distribution with parameters 1.47 and 42 200. Second line is the fraction of $R_k$ fitted to an exponential distribution with parameters 1 since $ER_k \approx 1$. Third line is a cross-plot of the $L_k$ and $P_k - \pi(A)L_k$. Here $cor(L_k, P_k - \pi(A)L_k) = -0.04$ and $cor(L_k^{-2}, (P_k - \pi(A)L_k)^2) = -0.06$. This is based on 1 953 intervals.

### Example 5, two modes

Assume the limiting density consists of two modes and $A$ and $B$ represent each of these modes. Assume further that the probability for moving between the two modes is independent of the state in the mode. Then the number of iterations needed in order to move from one mode to the other is exponentially distributed which correspond to a Weibull distribution with $k = 1$. Further, the recurrence interval, $L_k$, is the sum of two exponentially distributions which may be approximated with a Weibull distribution with $k \approx 1.5$, the same value as in the two other examples. If there were intermediate modes between $A$ and $B$ such that the recurrence interval was a sum of more than two exponential distributed variables, the recurrence interval would be less heavy tailed.

### Example 6, Cauchy distribution

Assume the limiting distribution is a Cauchy distribution, the Markov chain is a random walk with Gaussian distributed step length and the two subsets of the state space are $A = \{x|\ x_1 < -a\}$ and $B = \{x|\ x_1 > a\}$. In this case $H_{A,B}$ increases when $a$ increases. This implies that the Markov chain enters the subset $A$ less often than proportional with the probability $\pi(A)$ and this is compensated by staying longer in the area in and close to $A$ when $a$ increases. This will give very heavy tailed distribution that may be approximated by a Weibull distribution with $k < 1$, i.e. more heavy tailed than the exponential distribution. If we had assumed a Gaussian limiting distribution or a Cauchy distribution in the random walk, then $H_{A,B}$ had decreased and we would not have a heavy tail.

The last example shows that the length of the recurrence intervals may have more heavy tailed distribution than the exponential. But it is necessary with a quite extreme example in order to obtain this.

## 7. Closing remarks

This paper proposes a new type of recurrence and a function $H_{A,B}$ that gives us a better understanding of the mixing of Markov chains. Examples illustrate that the new recurrence intervals typically have length from a distribution with heavy tails. However, the length of the intervals was fitted with a Weibull distribution in three examples with $k$ approximately 1.5 which gives a distribution with less heavy tail than the exponential distribution. It is necessary with a quite extreme example in order to get heavier tails than in the exponential distribution. Heavy tails in the distribution of the length of the recurrence intervals make it necessary with more samples in order to be sure to sample the state space representatively.

We also give a bound on the variance the estimate on $\pi(A)$ after n iterations. We show that variance is proportional with $M_{A,B}$, the expected length of the recurrence intervals. Therefore, we should tune the acceptance rate and other parameters in the Markov chain in order to minimize $M_{A,B}$ and $H_{A,B}$. In some cases this implies choosing an acceptance rate that is far smaller than the interval (0.1,0.6) recommended in the literature. We find examples where it is optimal to have a much smaller acceptance rate than what is generally recommended in the literature and also examples where the optimal acceptance rate vanishes in the limit.

# Appendix

## Proof of the Proposition

When the proposal function is $\pi(A)$ then the probability for jumping from $A$ to $B$ in exactly n iterations is $Q_{A,B,n} = \pi(A)(1-\pi(A))^{n-1}$. Then the expected number of iterations needed in order to move from $B$ to $A$ is $\sum_{i=1} i\, \pi(A)(1-\pi(A))^{i-1} = \frac{1}{\pi(A)}$. Then a move from $A$ to $B$ followed by a move from $B$ to $A$ has the expected number of iterations $M_{A,B} = \frac{1}{\pi(A)} + \frac{1}{\pi(B)}$. $\square$

## Proof of the Theorem

The assumption in the Theorem ensures that the Markov chain has the same properties in all the intervals $I_j$. Then we have

$$\begin{aligned}
\frac{1}{\pi^2(A)} VAR(\widehat{\pi_n(A)}) &= VAR\left(\sum_{j=1}^{k} R_j \frac{L_j}{\sum_{s=1}^{k} L_s}\right) = VAR\left(\sum_{j=1}^{k} \frac{P_j}{\pi(A) \sum_{s=1}^{k} L_s}\right) \\
&= E\left(\sum_{j=1}^{k} \frac{P_j}{\pi(A) \sum_{s=1}^{k} L_s} - 1\right)^2 = E\left(\frac{1}{\sum_{s=1}^{k} L_s}\right)^2 \left(\sum_{j=1}^{k} \left(\frac{P_j}{\pi(A)} - L_j\right)\right)^2 \\
&\leq E\left(\frac{1}{\sum_{s=1}^{k} L_s}\right)^2 E\left(\sum_{j=1}^{k} \left(\frac{P_j}{\pi(A)} - L_j\right)\right)^2 \\
&= E\left(\frac{1}{\sum_{s=1}^{k} L_s}\right)^2 VAR\left(\sum_{j=1}^{k} \left(\frac{P_j}{\pi(A)} - L_j\right)\right)
\end{aligned}$$

We have used the first bound on the covariance in the inequality. The bound on the second covariance makes it possible to continue the calculation:

$$VAR\left(\sum_{j=1}^{k} \left(\frac{P_j}{\pi(A)} - L_j\right)\right)$$

$$= \sum_{j=1}^{k} VAR\left(\frac{P_j}{\pi(A)} - L_j\right) + \sum_{j=1}^{k} \sum_{i=1, i \neq j}^{k} COV\left(\frac{P_j}{\pi(A)} - L_j, \frac{P_i}{\pi(A)} - L_i\right)$$

$$\leq k\, VAR\left(\frac{P_1}{\pi(A)} - L_1\right) + \sum_{j=1}^{k} \sum_{i=1, i \neq j}^{k} c\, 2^{-|i-j|-1} VAR\left(\frac{P_1}{\pi(A)} - L_1\right)$$

$$\leq k\,(1+c) VAR\left(\frac{P_1}{\pi(A)} - L_1\right).$$

Since $\frac{P_j}{\pi(A)} - L_j = (R_j - 1)L_j$, we write these two bounds slightly differently in order to show the dependence on $\pi(A)$ and the independence of the scale of $L_i$.

$$\frac{1}{\pi^2(A)} VAR(\widehat{\pi_n(A)}) \leq E\left(\frac{k\, M_{A,B}}{\sum_{s=1}^{k} L_s}\right)^2 VAR\left(\sum_{j=1}^{k}(R_j - 1)\frac{L_j}{kM_{A,B}}\right)$$

and

$$\frac{1}{\pi^2(A)} VAR(\widehat{\pi_n(A)}) \leq \frac{1+c}{k} E\left(\frac{k\, M_{A,B}}{\sum_{s=1}^{k} L_s}\right)^2 VAR\left((R_1 - 1)\frac{L_1}{M_{A,B}}\right)$$

This proves the Theorem. □

### Example where $H_{A,B} < 1$

Assume there are 2n+1 states where $A = \{s_1, s_3, \ldots, s_{2n+1}\}$ and $B = \{s_2, s_4, \ldots, s_{2n}\}$ and all states are equally likely in the limiting density. Use a Metropolis-Hasting simulation algorithm that from state $s_i$ proposes state $s_{i-1}$ and $s_{i+1}$ with probability 0.5 each where we use the cyclic definition $s_0 = s_{2n+2}$. This means that the Markov chain changes between subset $A$ and $B$ in each iteration except when the chain is in the two neighboring states $s_1, s_{2n+1} \in A$. Then $M_{A,B} = \frac{4n+5}{2n+1}$ and $H_{A,B} = \frac{n(n+1)(4n+5)}{(2n+1)^3} \approx 0.5$ for $n$ large. Notice that in this example, the two subsets $A$ and $B$ are as close together as possible instead of far apart. The example shows that the subsets $A$ and $B$ must be far apart in order for $H_{A,B}$ to give valuable information about the mixing of the chain and not only between subsets $A$ and $B$.

## Acknowledgement

The author thanks Marit Holden for simulating the climate model and Arnoldo Frigessi for several good advice.

## References

Aldrin, M., Holden, M., Guttorp, P., Skeied, R.B., Myhre, G. and Berntsen, T.K. (2012) Bayesian estimation of climate sensitivity based on a simple climate model fitted to observations of hemispheric temperatures and global ocean heat content, *Environmetrics,* published online 24. February 2012, DOI: 10.1002/env.2140

Cowles, M. K. and Carlin, B.P., (1996) Markov Chain Monte Carlo Convergence Diagnostics: A Comparative Review. *J. Amer. Stat. Assoc.,* 91 (434), 883-904. June 1996.

Diaconis, P. and Stroock, D. Geometric Bounds for Eigenvalues of Markov Chains. (1991) Ann. Appl. Prob. 1(1) 36-61.

Gelman, A., Roberts, G.O., and Gilks, W.R., (1996) Efficient Metropolis jumping rules. In Byesian Statistics 5, ed. J. Bernardo et al. 599-607. Oxford University Press.

Giordani, P. and Kohn, R. (2010) Adaptive independent Metropolis-Hastings by fast estimation of mixtures of normals. *Journal of Computational and Graphical Statistics*, *19(2), 243-259.*

Holden, L. (1998) Geometric convergence of the Metropolis-Hastings Simulation Algorithm. *Statistics and Probability Letters.* 39, (4), 371-377.

Meyn S. P. and Tweedie, R. L. (1993) Markov Chains and Stochastic Stability, Springer-Verlag.

Neal, P. and Roberts, G. (2006) Optimal Scaling for Partially Updating MCMC Algorithms, *The Annals of Applied Probability*, 16 (2) 475-515.

Roberts, G.O., Gelman, A. and Gilks, W.R. (1997) Weak convergence and optimal scaling of Random Walk Metropolis algorithms. Ann. Appl. Prob. 7 110-120.

Roberts, G.O. and Tweedie R.L. (1999). Bounds on regeneration times and convergence rates for Markov chains. *Stoch. Proc. Appl*. 80 211-229. See also corrigendum Stoch. Proc Appl. 91 (2001) 337-338.

Rosenthal, G.O. and Rosenthal P. (1998). Optimal scaling of discrete approximations to Langevin diffusions. J. Roy. Stat. Soc. B 60, 255-268.

Rosenthal, G.O. and Rosenthal P. (2015). Spectral Bounds for Certain Two-Factor Non-Reversible MCMC Algorithms, Electronic communications in Probability, Vol 20, article 91.

Rosenthal, J. S. (2010), Optimal Proposal Distributions and Adaptive MCMC. Brooks, A. Gelman, G. Jones and X.-L. Meng (eds). Handbook of Markov Chain Monte Carlo. *Chapman &Hall/CRC*.

Skeie, R.B., Berntsen, T. , Aldrin, M., Holden, M. and Myhre, G. (2014) A lower and more constrained estimate of climate sensitivity using updated observations and detailed radiate forcing time series. *Earth Syst. Dynam*. 5, 139-175.